\documentclass[12pt]{article}

\usepackage[body={17.5cm, 21cm},right=2cm]{geometry}

\usepackage{color}
\usepackage{graphicx}
\usepackage{epsf}
\usepackage{graphicx,epsfig}
\usepackage{amsmath}
\usepackage{amssymb}
\usepackage{epsfig}
\usepackage{cite}
\usepackage{color,colordvi}
\newcommand{\be}{\begin{eqnarray}}
\newcommand{\ee}{\end{eqnarray}}
\newcommand{\bi}{\begin{itemize}}
\newcommand{\ei}{\end{itemize}}

\newcommand{\bx}{{\vec{x}}}
\newcommand{\bn}{{\vec{b}}}

\newcounter{hran}

\def\MSbar{\relax\ifmmode\overline{\rm MS}\else{$\overline{\rm MS}${ }}\fi}
\def\del{\partial}

\def\d{\rm d}

\def\d{{\rm d}}

\def\vq{\vec{q}}
\def\vx{\vec{x}}
\def\bb{{\vec{b}}}
\def\bx{{\vec{x}}}
\def\bk{{\vec{k}}}
\def\bnabla{{\bf \nabla}}
\def\bv{{\vec{v}}}
\def\t{\tau}

\def\r{\rho}

 \def\vx{\vec{ x}} 
\def\vk{\vec{k}}
\def\vq{\vec{q}}

\numberwithin{equation}{section}
\begin{document}
\vspace{5mm}
\vspace{0.5cm}
\begin{center}

\def\thefootnote{\fnsymbol{footnote}}

{\Large \bf 
Symmetries and Consistency Relations \\
\vspace{0.25cm}	
in the 	
Large Scale Structure of the Universe
}
\\[1.5cm]
{\large  A. Kehagias$^{a}$ and A. Riotto$^{b}$}
\\[0.5cm]

\vspace{.3cm}
{\normalsize {\it  $^{a}$ Physics Division, National Technical University of Athens, \\15780 Zografou Campus, Athens, Greece}}\\

\vspace{.3cm}
{\normalsize { \it $^{b}$ Department of Theoretical Physics and Center for Astroparticle Physics (CAP)\\ 24 quai E. Ansermet, CH-1211 Geneva 4, Switzerland}}\\

\vspace{.3cm}


\end{center}

\vspace{3cm}

\hrule \vspace{0.3cm}
{\small  \noindent \textbf{Abstract} \\[0.3cm]
\noindent 
We study the symmetries enjoyed by the Newtonian equations of motion of the non-relativistic dark matter fluid coupled to gravity which give rise to the phenomenon of gravitational instability. We also discuss some consistency relations  involving the soft limit of the $(n+1)$-correlator functions of matter and galaxy overdensities.

\vspace{0.5cm}  \hrule
\vskip 1cm

\def\thefootnote{\arabic{footnote}}
\setcounter{footnote}{0}


\baselineskip= 15pt

\newpage 

\section{Introduction}
\noindent
There is no doubt that symmetries play a crucial  role in high energy physics  \cite{coleman}. They allow, for instance,  to
derive Ward  identities among correlation functions which remain valid even after renormalization \cite{WT}. 
Symmetries are also relevant in the cosmological setting and  have been the subject of a recent and intense activity. They are particularly useful   in  characterizing  the properties of  the  cosmological perturbations  generated by an  inflationary stage \cite{lrreview}.  During inflation  the de Sitter isometry group acts  as conformal group  on $\mathbb{R}^3$ when the fluctuations are on super-Hubble scales. In such a  regime,  the SO(1,4) isometry
of the de Sitter background is realized as conformal symmetry of the flat $\mathbb{R}^3$ sections  and   correlators are constrained by conformal invariance \cite{antoniadis,creminelli1,us1,us2}. This  applies in the case in which the cosmological perturbations are generated 
by light scalar fields other than the inflaton (the field that drives inflation). In the opposite case in which the inflationary perturbations
originate from  only one degree of freedom,   conformal consistency relations among the inflationary correlators
 have also been recently  investigated  \cite{creminelli2,hui,baumann1,baumann2}. The fluctuations in single-field inflation  are Goldstone bosons of a spontaneously broken dilation symmetry.  Being non-linearly realized, the broken symmetry is
still respected in Ward identities and leads to a relation between the variation of the $n$-point function of the comoving curvature perturbation $\zeta$ under dilation and  the squeezed limit of the $(n + 1)$-point function \cite{mal,cz}. These identities will be extremely useful in discriminating among the various mechanisms for the generation of the cosmological perturbations. For instance, the detection of a sizable primordial three-point correlator  in the squeezed limit would rule out all single-field  models where inflation is
driven by a single scalar field with canonical kinetic energy and an initial Bunch-Davies vacuum. 

When perturbations re-enter the horizon, they provide the seeds for the the large scale structure of the universe which grows  via the gravitational instability \cite{sb}. At early epochs, the growth of the density perturbations can be described by linear perturbation theory and the perturbation Fourier modes evolve independently from one another, thus conserving the statistical properties of the primordial perturbations. When the perturbations become nonlinear, the coupling between the different Fourier modes become relevant, inducing nontrivial correlations that modify the statistical properties of the cosmological fields. At intermediate quasi-linear scales the evolution of matter may be described analytically by extending the standard perturbation theory \cite{sb}, where one defines a series solution to the fluid
equations in powers of the initial density field. The $n$-th order term of the series for the density contrast grows as the $n$-th
power of the scale factor $a$ (for a pressureless fluid), thus  affecting its convergence properties. 

The need for improving theoretical predictions for the next generation of very large galaxy
surveys has spurred many efforts to go beyond the standard perturbation theory. For instance, the renormalized perturbation theory \cite{rpt} reorganizes the perturbation expansion in terms of
different fundamental objects, the so-called non-linear propagator and non-linear vertices, to improve
the convergence. The renormalization group method \cite{rg} represents an alternative possibility where truncating the renormalization equation  at the level of some $n$-point  correlator leads to a solution that corresponds to the summation of an infinite class of
perturbative corrections. Other methods have been proposed in Refs. \cite{val,mat,rb,effective}.

The goal of this paper is to investigate the underlying symmetries of the Newtonian equations of motion which 
describe the gravitational instability and their consequences.  Consider the non-relativistic fluid equations in 
the presence of gravity
 \be
 &&\frac{\partial \delta(\bx,\t)}{\partial \t}+\bnabla(1+\delta(\bx,\t))\bv(\bx,\t)=0\label{fl1},\\
 && \frac{\partial \bv(\bx,\t)}{\partial \t}+{\cal{H}}(\t)\bv(\vx,\t)
 +[\bv(\vx,\t)\cdot \bnabla]\bv(\bx,\t)=\-\bnabla\Phi(\bx,\t),\label{fl2}\\
 &&
 \nabla^2\Phi(\bx,\t)=\frac{3}{2} \Omega{\cal{H}}^2(\tau)\delta(\bx,\tau), \label{fl3}
 \ee
where we have  denoted by $\bx$  the comoving spatial coordinates, $\tau=\int \d t/a$  the conformal time, $a$ the scale factor
in the FRW metric, 
${\cal{H}}=\d\ln(a)/\d\t$  the conformal expansion rate and, in addition, 
 $\delta(\bx,\t)=(\rho(\bx,\t)/\bar{\rho}-1)$ is the overdensity over the mean matter density density $\bar{\rho}$, 
$\bv(\bx,\t)$ is the velocity of the fluid away from  the Hubble flow and $\Phi(\bx,\t)$ is the gravitational 
potential due to density 
fluctuations. Finally $
\Omega=8\pi G\bar \rho a^2/3{\cal{H}}^2$
is the density parameter. 

We are interested in exploring possible symmetries of these equations (which are expected to correctly describe the gravitational instability when  vorticity and multi streaming are not present). In other words, we are looking for 
which  (possibly) non-linearly realized transformations of the coordinates and fields 
 
\begin{eqnarray}
\tau&\rightarrow & \tau',\nonumber\\
\bx&\rightarrow & \bx',\nonumber\\
\delta(\bx,\t)&\rightarrow &Z_\delta(\bx',\t')\delta(\bx',\t')+\xi_\delta(\bx',\t'),\nonumber\\
\bv(\bx,\t)& \rightarrow &Z_\bv(\bx',\t')\bv(\bx',\t')+\xi_\bv(\bx',\t'),\nonumber\\
\Phi(\bx,\t)&\rightarrow &Z_\Phi(\bx',\t')\Phi(\bx',\t')+\xi_\Phi(\bx',\t'),\nonumber
\end{eqnarray}
the Newtonian equations (\ref{fl1}-\ref{fl3}) describing the gravitational instability are invariant. 
The symmetries of the equations depend on the explicit form of ${\cal{H}}(\t)$. We will explore 
two particular cases, a namely matter-dominated era  and $\Lambda$CDM.
 
 Very much similar to what happens in quantum field theory, these symmetries lead to relations among the 
 correlation functions
which are valid at any order in perturbation theory.  As such,  they might be represent useful
consistency checks of the various analytical approaches and possibly  testable by observations.

One example of such a symmetry and its utility is represented by the Galilean symmetry: being the Newtonian
equations those of a classical non-relativistic  field theory, they are invariant under Galilean transformations. 
This invariance is indeed the underlying reason  for the cancellation of the leading  infrared divergences in the
computation of the matter  power spectrum observed in various perturbative schemes to arbitrary number of loops 
\cite{jb,scocgal,scocself}.

In fact, we will show that the Galilean transformations are only a special case of a more  general set of  transformations enjoyed by the non-relativistic fluid equations coupled to gravity. This general set contains also the  acceleration transformations where in the new system of coordinates the observer is uniformly accelerated. Furthermore, if the universe is matter-dominated,  the gravitational instability equations are invariant also under a Lifshitz scaling  symmetry with a generic exponent $z$ \cite{pd}.  The discussion of these symmetries will be performed in section 2 under the assumption that the universe is matter-dominated, while section 3 is devoted to the case in which the universe contains a non vanishing cosmological constant.  A more elegant discussion of the symmetries will be given in section 4 based on the derivation of the four-dimensional non-relativistic  fluid equations
 from the dimensional reduction of a five-dimensional scalar field theory. 
 Finally, section 5 will contain a discussion of the consistency relations which may obtained on the 
 correlators from the underlying symmetries.

\section{Symmetries during matter-domination} 
Let us start from the simplest possibility,  a matter-dominated universe. In such a case the scale 
factor $a$ scales like  $\t^2$ and  we have ${\cal{H}}=2/\t$. We will also assume  $\Omega=1$. Under
these assumptions, 
 the fluid equations Eqs. (\ref{fl1}-\ref{fl3}) are obviously invariant under  
 SO(3) space rotations as well as under the Galilean boosts

 \begin{align}
 \t'=\t,&~~~\bx'=\bx+\vec{u} \, \t,\label{gal1}\\
 \delta'(\bx,\t)&=\delta(\bx',\t'),\label{d1}\\
 \bv'(\bx,\t)&=\bv(\bx',\t')-\vec{u},\label{v1}\\
 \Phi'(\bx,\t)&=\Phi(\bx',\t')-\frac{2}{\t} \vec{u}\cdot\bx, \label{gf1}
 \end{align}
 where $\vec{u}$ is a constant three-dimensional vector.
Indeed, since 
\be
\frac{\partial}{\partial \t}=\frac{\partial}{\partial \t'}+\vec{u}\cdot \bnabla'\, , ~~~\bnabla=\bnabla', 
\ee
we have, for example for Eq. (\ref{fl2}),
\begin{align}
0&=\frac{\partial \bv'(\bx,\t)}{\partial \t}+{\cal{H}}(\t)\bv'(\vx,\t)
 +[\bv'(\vx,\t)\cdot \bnabla]\bv'(\bx,\t)-\bnabla\Phi'(\bx,\t)\nonumber \\
 &=\frac{\partial \bv(\bx',\t')}{\partial \t'}+{\cal{H}}(\t')\bv(\vx',\t')-
{\cal{H}}(\t)\vec{u}
 +[\bv(\vx',\t')\cdot \bnabla']\bv(\bx',\t')-\bnabla\Phi'(\bx,\t)\nonumber \\
 &=-
{\cal{H}}(\t)\vec{u}+\bnabla'\Phi(\bx',\t')-\bnabla\Phi'(\bx,\t).
\end{align}
Therefore, Eq. (\ref{fl2}) is invariant if 
\be
\bnabla\Phi'(\bx,\t)=\bnabla'\Phi(\bx',\t')-{\cal{H}}(\t)\vec{u}
\ee
or 
\be
\Phi'(\bx,\t)=\Phi(\bx',\t')-{\cal{H}}(\t)\,\vec{u}\cdot \bx. \label{gf11}
\ee
This is just Eq. (\ref{gf1}) for ${\cal{H}}=2/\t$. 
Using the transformation (\ref{gf11}) into (\ref{fl3}), we get
\begin{align}
0&= \nabla^2\Phi'(\bx,\t)-\frac{3}{2} \Omega{\cal{H}}^2(\tau)\delta'(\bx,\tau)=
\nabla'^2\Phi(\bx',\t')-\frac{3}{2} \Omega{\cal{H}}^2(\tau)\delta'(\bx,\tau)\nonumber \\
&=\frac{3}{2} \Omega{\cal{H}}(\tau')^2\delta(\bx',\tau')-\frac{3}{2} \Omega{\cal{H}}^2(\tau)\delta'(\bx,\tau),
\end{align}
from which the transformation (\ref{d1}) follows immediately. Finally, a simple inspection of Eq.(\ref{fl1})
shows that it is also invariant under (\ref{d1}-\ref{gf1}), thus recovering the well-known 
 invariance of the fluid equations under Galilean transformations. 
In the infinitesimal form, the transformations (\ref{gal1}-\ref{gf1}) take the form 
(with $\delta \vec{u}=\vec{b}$)

 \be
 &&\delta_{\rm g} \t=0,\, ~~\delta_{\rm g}\bx=\bb\, \t,\\
 &&\delta_{\rm g}  \delta(\bx,\t)=\t\, \bb\cdot\bnabla \delta(\bx,\t),\\
 &&\delta_{\rm g}\bv(\bx,\t)=\t\, \bb\cdot\bnabla \bv(\bx,\t)-\bb,\\
 &&\delta_{\rm g}\Phi(\bx,\t)=\t\, \bb\cdot\bnabla\Phi(\bx,\t)-\frac{2}{\t}\bb\cdot \bx.
 \ee
The Newtonian equations are also 
 invariant under the acceleration transformations
 \begin{align}
 \t'=\t,&~~~\bx'=\bx+\frac{1}{2}\vec{a} \t^2,\label{ac1}\\
 \delta'(\bx,\t)&=\delta(\bx',\t'),\label{da1}\\
 \bv'(\bx,\t)&=\bv(\bx',\t')-\vec{a} \t, \label{va1}\\
 \Phi'(\bx,\t)&=\Phi(\bx',\t')-3 \vec{a}\cdot\bx, \label{gfa1}
 \end{align}
 where $\vec{a}$ is a constant three-dimensional vector.
 Indeed, using 
 \be
\frac{\partial}{\partial \t}=\frac{\partial}{\partial \t'}+\t\, \vec{a}\cdot \bnabla', ~~~\bnabla=\bnabla',
\ee
we get 
\begin{align}
0&=\frac{\partial \bv'(\bx,\t)}{\partial \t}+{\cal{H}}(\t)\bv'(\vx,\t)
 +[\bv'(\vx,\t)\cdot \bnabla]\bv'(\bx,\t)-\bnabla\Phi'(\bx,\t)\nonumber \\
 &=\frac{\partial \bv(\bx',\t')}{\partial \t'}-\vec{a}+{\cal{H}}(\t')\bv(\vx',\t')-{\cal{H}}(\t')\tau \vec{a}
 +[\bv(\vx',\t')\cdot \bnabla']\bv(\bx',\t')-\bnabla\Phi'(\bx,\t)\nonumber \\
 &=-\vec{a}-
{\cal{H}}(\t')\vec{a} \t+\bnabla'\Phi(\bx',\t')-\bnabla\Phi'(\bx,\t)
\end{align}
Therefore Eq. (\ref{fl2}) is invariant if 
\be
\bnabla\Phi'(\bx,\t)=\bnabla'\Phi(\bx',\t')-\vec{a} -{\cal{H}}\vec{a}\t
\ee
or 
\be
\Phi'(\bx,\t)=\Phi(\bx',\t')-\vec{a}\cdot \bx-\t {\cal{H}}\vec{a}\cdot \bx. \label{gf2}
\ee
Again, for ${\cal{H}}=2/\t$, Eq. (\ref{va1}) follows. From the Poisson equation we get 
\begin{align}
0&= \nabla^2\Phi'(\bx,\t)-\frac{3}{2} \Omega{\cal{H}}^2(\tau)\delta'(\bx,\tau)=
\nabla'^2\Phi(\bx',\t')-\frac{3}{2} \Omega{\cal{H}}^2(\tau)\delta'(\bx,\tau)\nonumber \\
&=\frac{3}{2} \Omega{\cal{H}}^2(\tau')\delta(\bx',\tau')-\frac{3}{2} \Omega{\cal{H}}^2(\tau)\delta'(\bx,\tau),
\end{align}
so that $\delta(\bx,\t)$ is an invariant scalar as Eq. (\ref{da1}) shows. Finally, using Eq. (\ref{da1}-\ref{gfa1}), 
we find that (\ref{fl1}) is also invariant.
The corresponding infinitesimal form of the acceleration transformation is (with $\delta \vec{a}=\vec{b}$)
 \be
 &&\delta_{\rm a} \t=0\, ~~\delta_{\rm a}\bx=\frac{1}{2}\vec{b}\, \t^2, \label{trac}\\
 &&\delta_{\rm a}\delta(\bx,\t)=\frac{1}{2}\t^2\vec{b}\cdot \bnabla\delta(\bx,\t),\label{acd} \\
 &&\delta_{\rm a}\bv(\bx,\t)=\t \vec{b}\cdot\bnabla \bv( \bx,\t)-\vec{b} \t,\label{acv}\\
 &&\delta_{\rm a}\Phi(\bx,\t)=\frac{1}{2}\t^2\, \vec{b}\cdot \bnabla \Phi(\bx,\t)-3\vec{b}\cdot \bx. \label{acf}
 \ee
 In fact, Galilean  and  acceleration transformations 
are special cases of a most general transformation of the non-relativistic
 fluid equations
 \begin{align}
 \t'=\t,&~~~\bx'=\bx+\vec{n}(\t),\label{ag1}\\
 \delta'(\bx,\t)&=\delta(\bx',\t'),\label{dg1}\\
 \bv'(\bx,\t)&=\bv(\bx',\t')-\dot{\vec{n}}(\t), \label{vg1}\\
 \Phi'(\bx,\t)&=\Phi(\bx',\t')-\left(\ddot{\vec{n}}(\t)+\frac{2}{\t} \dot{\vec{n}}(\t)\right)\cdot \bx. \label{gfg1}
 \end{align}
 Note that we are using the dot  to denote differentiation with respect to conformal time $\t$. In 
 order to prove the invariance of the fluid equations under the transformations (\ref{ag1}-\ref{gfg1}), we note that 
 \be
 \label{dg}
\frac{\partial}{\partial \t}=\frac{\partial}{\partial \t'}+\dot{\vec{n}}\cdot \bnabla'\, , ~~~\bnabla=\bnabla',
\ee
which means that the operator
\be
D_\t=\frac{\partial}{\partial \t}+\bv(\bx,\t)\cdot \bnabla
\ee
is invariant under the transformation (\ref{ag1}).  From Eq. (\ref{fl2}) we therefore get 
\begin{align}
\label{gen1}
0&=\frac{\partial \bv'(\bx,\t)}{\partial \t}+{\cal{H}}(\t)\bv'(\bx,\t)
 +[\bv'(\bx,\t)\cdot \bnabla]\bv'(\bx,\t)-\bnabla\Phi'(\bx,\t)\nonumber \\
 &=\frac{\partial \bv(\bx',\t')}{\partial \t'}-\ddot{\vec{n}}+{\cal{H}}(\t')\bv(\bx',\t')
 -{\cal{H}}(\t')\dot{\vec{n}}
 +[\bv(\vx',\t')\cdot \bnabla']\bv(\bx',\t')-\bnabla\Phi'(\bx,\t)\nonumber \\
 &=-\ddot{\vec{n}}-
{\cal{H}}(\t')\dot{\vec{n}}+\bnabla'\Phi(\bx',\t')-\bnabla\Phi'(\bx,\t), 
\end{align}
 from which we deduce the transformation 
\be
\bnabla\Phi'(\bx,\t)=\bnabla'\Phi(\bx',\t')+\ddot{\vec{n}}+
{\cal{H}}(\t')\dot{\vec{n}}
\ee
or
\be
\Phi'(\bx,\t)=\Phi(\bx',\t')-\left(\ddot{\vec{n}}+
{\cal{H}}(\t')\dot{\vec{n}}\right)\cdot \bx.
\ee
It  is a straightforward  exercise to check the invariance  of Eqs. (\ref{fl1}) and (\ref{fl3})  under 
the general transformations (\ref{dg1}),(\ref{vg1}), and (\ref{gfg1}). 
Their infinitesimal form are given by (with $\delta\vec{n}=\bn$) 
 \be
 &&\delta_{\rm gen} \t=0,\, ~~\delta_{\rm gen}\bx=\bn(\t),\\
 &&\delta_{\rm gen}  \delta(\bx,\t)=\bn(\t)\cdot \bnabla\delta(\bx,\t),\\
 &&\delta_{\rm gen}\bv(\bx,\t)= \vec{b}(\t)\cdot\bnabla \bv( \bx,\t)-\dot{\bn}(\t),\\
 &&\delta_{\rm gen}\Phi(\bx,\t)= 
 {\vec{b}}(\t)\cdot \bnabla \Phi(\bx,\t)-
 \left(\ddot{\bn}(\t)+\frac{2}{\t}\dot{\bn}(\t)\right)\cdot \bx.
 \ee
Fluid equations during the matter-dominated period are also invariant under Lifshitz scalings of the form 
 
\begin{align}
 \t'=\lambda^z\t,&~~~\bx'=\lambda \bx,\label{as1}\\
 \delta'(\bx,\t)&=\delta(\bx',\t'),\label{ds1}\\
 \bv'(\bx,\t)&=\lambda^{z-1}\bv(\bx',\t'), \label{vs1}\\
 \Phi'(\bx,\t)&=\lambda^{2(z-1)}\Phi(\bx',\t') \label{gfs1}
 \end{align}
 for a generic Lifshitz weight $z$. 
Indeed,   using 
 \be
\frac{\partial}{\partial \t}=\lambda^z\frac{\partial}{\partial \t'}\, , ~~~\bnabla=\lambda \bnabla',
\ee
we get 
\begin{align}
0&=\frac{\partial \bv'(\bx,\t)}{\partial \t}-\vec{a}+{\cal{H}}(\t)\bv'(\vx,\t)
 +[\bv'(\bx,\t)\cdot \bnabla]\bv'(\bx,\t)-\bnabla\Phi'(\bx,\t)\nonumber \\
 &=\lambda^{2z-1}\frac{\partial \bv(\bx',\t')}{\partial \t'}+
 \lambda^{z-1}{\cal{H}}(\t'\lambda^{-z})\bv(\bx',\t')
 +\lambda^{2z-1}[\bv(\bx',\t')\cdot \bnabla']\bv(\bx',\t')- \bnabla\Phi'(\bx,\t).
 \end{align}
 We see immediately that only when ${\cal{H}}$ scales like $1/\t$, the above equation is covariant. In this case
 we get 
 \begin{align}
 0&=\lambda^{2z-1}\left(\frac{\partial \bv(\bx',\t')}{\partial \t'}-\frac{2}{\tau'}\bv(\bx',\t')
 +[\bv(\bx',\t')\cdot \bnabla']\bv(\bx',\t')\right)- \bnabla\Phi'(\bx,\t)\nonumber \\
&=\lambda^{2z-1}\bnabla'\Phi(\bx',\t')-\bnabla\Phi'(\bx,\t),
\end{align}
that is
\be
\bnabla\Phi'(\bx,\t)=\lambda^{2z-1}\bnabla'\Phi(\bx',\t')
\ee
or
\be
\Phi'(\bx,\t)=\lambda^{2z-2}\Phi(\bx',\t').
\ee
Using this transformation in Poisson equation we find 
\begin{align}
0&= \nabla^2\Phi'(\bx,\t)+\frac{3}{\tau^2} \Omega\delta'(\bx,\tau)=
\lambda^{2z}\nabla'^2\Phi(\bx',\t')-\lambda^z\frac{3}{\t'^2}\delta'(\bx,\tau)\nonumber \\
&=-\lambda^{2z}\frac{3}{\t'^2} \Omega\delta(\bx',\tau')-\lambda^{2z}\frac{3}{\tau'^2} \delta'(\bx,\tau),
\end{align}
from which the transformation (\ref{ds1}) immediately follows. Then, it is straightforward to check the invariance 
of Eq. (\ref{fl1}) under the transformations (\ref{ds1}),(\ref{vs1}) and (\ref{gfs1}).
The infinitesimal forms of the Lifshitz  scalings read (with $\lambda$ to be intended as an infinitesimal parameter)
 \be
 &&\delta_{\rm l} \t=z\lambda \t\, , ~~~\delta_{\rm l} \bx=\lambda \bx,\\
 &&\delta_{\rm l} \delta(\bx,\t)=\lambda \, \bx\cdot\bnabla \delta(\bx,\t)+z\,\lambda \,\t\frac{\partial}{\partial \t}\delta(\bx,t),\\
 && \delta_{\rm l}\bv(\bx,\t)=\lambda \, \bx\cdot\bnabla \bv(\bx,\t)+z\lambda \t\frac{\partial}{\partial \t}\bv(\bx,t)
 +(z-1)\lambda \bv(\bx,\t),\\\, 
 &&\delta_{\rm l}\Phi(\bx,\t)=\lambda \, \bx\cdot\bnabla \Phi(\bx,\t)+
 z \lambda \t\frac{\partial}{\partial \t} \Phi(\bx,\t)+2(z-1)\lambda \Phi(\bx,\t).
\ee
Notice that in the linear regime, when the gravitational potential $\Phi(\bx,\t)$ does not evolve as function of time, $\Phi(\bx,\t)$ has a Lifshitz weight equal to $2(z-1)$. Going beyond the linear order and writing $\Phi(\bx,\t)=\sum_{n\geq 0} \Phi^{(n)}(\bx)\tau^{2n}$, one finds that
$\Phi^{(n)}(\bx)$ has Lifshitz weight $(2n+1)z-2$.

 \section{Symmetries in $\Lambda$CDM} 
 In this section we investigate the symmetries if the universe, besides a matter component, is characterized by 
 the presence of a non vanishing
 cosmological constant. Having learnt that in the matter-dominated case there is a set of transformations which contain Galilean and acceleration transformations as particular case, 
 let us consider the  more  general  set of transformations
 \be
 \t'=\t, ~~~\bx'=\bx+{\vec{n}}(T),
 \label{gen}
 \ee
  where
   \be
   T(\tau)=\frac{1}{a(\t)}\int^\t \d\eta \,a(\eta).
    \ee
Following the same steps of the previous section, 
it can be checked that the non-relativistic fluid equations  are invariant under 
\begin{align}
\delta'(\bx,\t)&=\delta(\bx',\t'),\label{g1}\\
\bv'(\bx,\t)&=\bv(\bx',\t')-\dot{\vec{n}}(T),\label{g2}\\
\Phi'(\bx,\t)&=\Phi(\bx',\t')+\Big{(}{\cal{H}}\dot{\vec{n}}(T)+\ddot{\vec{n}}(T)\Big{)}\cdot \bx,\label{g3}
\end{align}
where again we remind the reader that the dot denotes  differentiation with respect to conformal time $\t$. 
The proof proceeds as in Eq. (\ref{gen1}) after using (\ref{dg}). These transformation include  
the Galilean boosts for which

\be
{\vec{n}}(T)=T \vec{u}, \,\,\,
\frac{\partial T}{\partial \t}=1-{\cal{H}} T, ~~~~~\frac{\partial^2T}{\partial \t^2}=-\dot{{\cal{H}}}T-
{\cal{H}}(1-{\cal{H}} T).
\ee
Eqs. (\ref{g1}-\ref{g3})  become
\begin{align}
\delta'(\bx,\t)&=\delta(\bx',\t'),\label{g11}\\
\bv'(\bx,\t)&=\bv(\bx',\t')-\vec{u}(1-{\cal{H}} T),\label{g12}\\
\Phi'(\bx,\t)&=\Phi(\bx',\t')-T\frac{\partial{\cal{H}}}{\partial \t}\vec{u}\cdot \bx. \label{g13}
\end{align}
Acceleration transformations are also included as a special case. It is enough to take
\be
{\vec{n}}(T)=\frac{1}{2}\vec{u}\,T^2, 
\ee
and Eqs. (\ref{g1}-\ref{g3}) become
\begin{align}
\delta'(\bx,\t)&=\delta(\bx',\t'),\label{g11}\\
\bv'(\bx,\t)&=\bv(\bx',\t')-\frac{1}{2}\vec{u}\, T(1-{\cal{H}} T),\label{g12}\\
\Phi'(\bx,\t)&=\Phi(\bx',\t')+\left((1-{\cal{H}}T)^2-T^2\frac{\partial{\cal{H}}}{\partial \t}\right)
\vec{u}\cdot \bx. \label{g13}
\end{align}
One can also easily verify that the Lifshitz scaling is no longer a symmetry of the non-relativistic equations when the universe is of the
$\Lambda$CDM type.

\section{Fluid dynamics by dimensional reduction and symmetries}
The purpose of this section is to show how the symmetries of the non-relativistic equations describing the gravitational instability may be though of as symmetries of higher dimensional theory on which dimensional reduction is applied \cite{Duval,pig,gibbons}. We are going to do it step by step, starting from a flat geometery and adding gravity and the expansion of the universe later on. 
Consider first the five-dimensional Bargmann spacetime with metric 
\be 
\d s^2=2 \d t \d\xi+\d \bx ^2, \label{xit}
\ee
where $\xi $ is a null direction. The dynamics of a free complex scalar field $\phi$ in such a space-time is described by the Lagrangian
\be
{\cal{L}}_5=-\partial_\mu \phi^\dagger \partial^\mu \phi, \label{phiphi}
\ee
which can explicitly be written as
\be
 {\cal{L}}_5=-\del_t \phi^\dagger \del_\xi \phi-\del_\xi \phi^\dagger \del_t \phi- |\nabla \phi|^2. \label{pps}
\ee 
For 
$\phi(t,\xi,\bx)$  of the form
\be
\phi(t,\xi,\bx)=e^{im \xi} \psi(t,\bx) \label{phi}
\ee
the Lagrangian becomes
\be
 {\cal{L}}_5= im \left(\psi^\dagger \del_t \psi-\psi \del_t \psi^\dagger\right)-|\nabla \psi|^2. \label{L1}
\ee
The corresponding equation of motion for 
 $\psi(t,\bx)$ is 
\be
-\nabla^2 \psi(t,\bx)=2 i m\del_t\psi(t,\bx),
\ee
which is the well-known  Schr\'odinger equation.
The theory (\ref{L1}) is invariant under the full Schr\"odinger group ${\rm Sch(3)}$, that is  the symmetry group of a free non-relativistic 
theory. In general, ${\rm Sch(d)}$ is defined as the subgroup of SO(2,$d$+2) that leaves invariant the 
momentum along a null direction.  Note that ${\rm Sch(3)}$ contains Lifshitz scaling  symmetry with $z=2$,  as  one may directly verify. 

We may also express $\psi(t,\bx)$ as 
\be
\psi(t,\bx)=\sqrt{\frac{\rho(t,\bx)}{m}}e^{i m\omega(t,\bx)}, \label{fr}
\ee
where $\rho(t,\bx)$ and $\omega(t,\bx)$ are  real functions and they will play the role of the matter 
density and the velocity potential, respectively. 
The Lagrangian (\ref{pps})   written in terms of $\r(t,\bx)$ and $\omega(t,\bx)$
becomes
\be
 {\cal{L}}_5=-2m\rho\, \del_t \omega-\frac{1}{4 m\rho}(\nabla \rho)^2-m\rho\,(\nabla \omega)^2. \label{L2}
\ee
If we now take the limit  of the momentum $m$ along the $\xi$-direction to infinity,  $m\to \infty$,  we get 
\be
 {\cal{L}}_5=-2m \left(\rho\, \del_t \omega +\frac{1}{2}\rho\, (\nabla\omega)^2\right)+{\cal{O}}\left(\frac{1}{m}\right). \label{L22}
\ee
The theory described by the Lagrangian  (\ref{L22}) is invariant under the Galilei group
\be
t\to t'=t,  ~~~ \bx \to \bx '=\bx +\bv t,
\ee
if $\rho$ and $\omega$ transform as
\be
\rho\to \rho'=\rho, ~~~\omega\to \omega'=\omega-\bv\cdot\bx -\frac{1}{2}v^2 t.
\ee
Furthermore, the theory is invariant under the arbitrary Lifshitz scaling $t'=\lambda^z t$ and $\bx'=\lambda\bx$ if we assign the following weights
\be
\omega\to \lambda^{2-z} \omega \, , ~~~~\rho\to \lambda^{-5+z} \rho
\ee
to $\rho(t,\bx)$ and $\omega(t,\bx)$. Note that for finite $m$, the theory (\ref{L2}) is invariant under the full ${\rm Sch(3)}$ group which, however, 
is broken down to the  Galilei  group plus   arbitrary Lifshitz scalings in the $m\to \infty$ limit.  

Let us now  include gravity in the theory (\ref{phiphi}). For this, we perturb the flat metric (\ref{xit}) as 
\be
\d s^2=-2 \Phi(t,\bx)\d t^2+2\left(1+\Phi(t,\bx)\right)  \d t \d\xi+\left(1+\Phi(t,\bx)\right) \d\bx ^2,
\label{withgravity}
\ee
where $\Phi(t,\bx)$ is the Newtonian potential. Then the dynamics is described by 
the Lagrangian
\be
{\cal{L}}_5=\frac{M_*^3}{2}R-\partial_\mu \phi^\dagger \partial^\mu \phi, \label{lg}
\ee
where $M_*$ plays the role of the five-dimensional  reduced Planck mass. 
Replacing the ansatz  (\ref{phi},\ref{fr}) 
for 
the scalar $\phi(t,\bx)$  we have 
\be
{\cal{L}}_5=-\left(-im \left(\psi^\dagger \del_t \psi-\psi \del_t \psi^\dagger\right)+|\nabla \psi|^2\Phi
+m^2 \psi^\dagger\psi\Phi+\frac{1}{2} M_*^3(\nabla \Phi)^2\right)
\ee
and, after using the ansatz (\ref{fr}), the Lagrangian becomes in terms of 
 the fields  $\rho(t,\bx)$ and $\omega(t,\bx)$ 
\be
 {\cal{L}}_g=-\left(2m\rho\, \del_t \omega+\frac{1}{4m \rho}(\nabla \rho)^2+m\rho\,(\nabla \omega)^2+2 m \rho \Phi+
\frac{1}{2}  M_*^3(\nabla \Phi)^2\right).
\ee
Upon taking the  $m\to \infty$ limit,  we finally get 
\be
 {\cal{L}}_5=-2m\left\{\rho\left( \del_t \omega+\frac{1}{2}\,(\nabla \omega)^2+ \Phi\right)+
M_{\rm p}^2(\nabla \Phi)^2\right\} +{\cal{O}}\left(\frac{1}{m}\right), \label{L33}
\ee
which describes a fluid in gravitational field after the following identification 
\be
M_*^3=4 m M_{\rm p}^2,
\ee
where $M_{\rm p}$ is the four-dimensional reduced Planck mass. 
The term of order ${\cal{O}}(1/m)$ we are neglecting is $(\nabla \rho)^2/(m\rho)$ and therefore the Lagrangian we have derived is valid as long as $
|\nabla\rho/m\rho\nabla\omega|\ll 1$. 
Note that the theory (\ref{L33}) is invariant only under the Galilei group as the presence 
of the gravitational potential breaks the the arbitrary  Lifshitz scaling invariance. 

Finally, we  introduce the cosmological expansion. As we shall see, this is essential in restoring the arbitrary  
Lifshitz scaling invariance. 
Consider  the metric

\be
\d s^2=-2\Phi(t,\bx)\d t^2+2\left(1+\Phi(t,\bx)\right) \d t \d\xi+a^2(t)\left(1+\Phi(t,\bx)\right)\d\bx ^2.
\ee
Upon dimensional reduction with $\phi(t,\xi,\bx)$ of the form given in Eqs. (\ref{phi},\ref{fr})  
we obtain in the  $m\to \infty$ limit 
\be
S=\int \d t \d^3 \vx\, \left\{-2m \left\{a^3\rho \left( \del_t \omega+\frac{1}{2a^2}|\nabla \omega|^2+ \Phi\right)+
aM_{\rm  p}^2(\nabla \Phi)^2\right\} +{\cal{O}}\left(\frac{1}{m}\right)\right\}. \label{c5}
\ee
The equations of motions for $\rho(t,\bx)$, $\omega(t,\bx)$ and $\Phi(t,\bx)$ are then 
 precisely the Newtonian  equations describing the fluid gravitationally instability 
of a non-relativistic fluid with density $\rho$ and velocity $\bv(t,\bx)=\bnabla \omega(t,\bx)/a$, using cosmic time.

Let us try now to identify symmetries of the action (\ref{c5}). It can easily be checked that the following 
transformations
\begin{align}
 &\delta_{\rm gen} t=0\, ~~~\delta_{\rm gen}\vx=\vec{n}(t),\\
 &\delta_{\rm gen}\omega(t,\vx)=-
 a^2\frac{\partial \vec{n}(t)}{\partial t}\cdot \vx+\vec{n} \cdot \bnabla \omega,\\
 &\delta_{\rm gen}\Phi(t,\vx)=\frac{\partial}{\partial t}\left(a^2 \frac{\partial \vec{n}(t)}{\partial t}\right)\cdot \vx +\vec{n}\cdot \bnabla \Phi  \\
 &\delta_{\rm gen}\rho(t,\vx)=\vec{n}\cdot \bnabla \rho,
\end{align}
leave (\ref{c5}) invariant. Of course, the above transformation includes Galilean boosts ($\vec{n}=\vec{n}_0 \t$, 
$\vec{n}_0$ constant vector) as well as acceleration ($\vec{n}=\vec{n}_0 \t^2$). 
In addition, there is also a Lifshitz symmetry for the action (\ref{c5}). Assuming the rescalings
\be
&&t\to t'=\lambda_t t,\\
&& a\to a'=\lambda_a a,\\
&&\vx\to \vx'=\lambda \vx,\\
&& \omega\to \omega'=\lambda_\omega \omega,\\
&&\rho\to \rho'=\lambda_\rho \rho, \\
&& \Phi\to \Phi'=\lambda_\Phi \Phi,
\ee
and that the scale factor $a$ is a homogeneous function of time (so that $a\to a'=\lambda_a a$), 
it is easy to verify that (\ref{c5}) is invariant for
\be
\lambda_\omega=\lambda^{(z-1)/2}\, , ~~~\lambda_\Phi=\lambda^{2-2z}\, , ~~~
\lambda_\rho=\lambda^{5-5z}.\, , ~~~\lambda_a=\lambda^{(3z-5)/2},\, ~~~\lambda_t=\lambda^{(5z-5)/2} \label{lif}
\ee
We have taken  conformal time to change as $\t\rightarrow\t'=\lambda^z\t$, so that
 $\lambda_a=\lambda_t\lambda^{-z}$. Note that when $a$ is not a homogeneous function of time, as in the $\Lambda$CDM case form 
example, the Lifshitz scaling is not  a symmetry.

\section{Consistency relations and conclusions}
As the gravitational instability equations posses a set of symmetries, it is natural to ask what are their consequences. The popular choice for the initial conditions in cosmology is that of a (nearly) Gaussian random field which is statistically homogeneous and isotropic in space. The distribution in any one realization is therefore not expected to be satisfying, {\it e.g.},  the Lifshitz scaling we have discussed for the matter-dominated period. The symmetries should be intended to play a role only at the statistical level in the sense that the correlators of a given observable, for example the matter density contrast, should have the same statistical properties of its transform under a given symmetry. This means that the correlators should satisfy 
 appropriate Ward identities, reflecting  the invariance under the given symmetries. For example,
  for the $n$-point connected  correlators of the gravitational potential at equal time
 \be
 G_\Phi^{(n)}(\vx_1,\vx_2,\ldots,\vx_n,\t)=\Big<\Phi(\vx_1,\t),\Phi(\vx_2,\t)\cdots \Phi(\vx_n,\t)\Big>_c,
 \ee
 the Ward identity follows from
 \be
 0=\delta G_\Phi^{(n)}(\vx_1,\vx_2,\ldots,\vx_n,\t)= \Big<\delta\Phi(\vx_1,\t),\Phi(\vx_2,\t)\cdots \Phi(\vx_n,\t)\Big>_c+
 \mbox{permutations}. \label{ward}
\ee
For the Galilean and  acceleration  symmetry it is easy to see that Eq. (\ref{ward})  
 is satisfied for $G^{(n)}_c=G^{(n)}_c(\bx_{ij},\t)$ where $\bx_{ij}=(\bx_i-\bx_j)$. 
 In addition, rotational invariance imposes that $
G^{(n)}_c=G^{(n)}_c(x_{ij},\t)$, where $x_{ij}=\left|\bx_i-\bx_j\right|$. 
The Lifshitz scaling  symmetry on the other hand gives 
\be
\left(2n(z-1) +z \,\t \frac{\partial}{\partial\t}+ \sum_{i<j}\bx_{ij}\cdot\nabla_{ij} \right)G^{(n)}_\Phi(x_{ij},\t)=0.
\ee
For the two-point correlator one finds therefore that $G^{(2)}_c$ should satisfy
\be
\left(4(z-1) +z \,\t \frac{\partial}{\partial\t}+ \bx_{12}\cdot\bnabla_{12} \right)G^{(2)}_\Phi(x_{12},\t)=0.
\ee
Solving the above equation, we get that 
\be
G^{(2)}_\Phi(\bx_1,\bx_2,\t)=\frac{1}{x_{12}^{4(z-1)}}\,  {\cal F}_\Phi\left(\frac{\t}{x_{12}^z}\right).
\ee
where ${\cal F}_\Phi$ is a function of only $\t/x_{12}^z$. 
Similarly, for the three-point correlator we get the equation
\be
\left(6(z-1) +z \,\t \frac{\partial}{\partial\t}+ \bx_{12}\cdot\bnabla_{12}+\bx_{13}\cdot\bnabla_{13}+
\bx_{23}\cdot\bnabla_{23} \right)G^{(3)}_\Phi(\bx_1,\bx_2,\bx_3,\t)=0.
\ee
The solution of the above equation which is also invariant under the permutation $\bx_i\to \bx_j, ~(i,j=1,2,3)$ is  
\be
G^{(3)}_\Phi(\bx_1,\bx_2,\bx_3,\t)=\frac{1}{x_{12}^{2(z-1)}x_{13}^{2(z-1)}x_{23}^{2(z-1)}}
\, {\cal G}_\Phi\left(\frac{\t}{x_{12}^z},\frac{\t}{x_{13}^z},\frac{\t}{x_{23}^z} \right).
\ee
Similarly, one finds that the two-point connected correlator of the matter over density at equal time
must satisfy the following relation because of the  Lifshitz scaling  symmetry \cite{pd,sb}

\be
G^{(2)}_\delta(\bx_1,\bx_2,\t)=\Big<\delta(\vx_1,\t),\delta(\vx_2,\t)\Big>_c={\cal F}_\delta\left(\frac{\t}{x_{12}^z}\right).
\ee
 If one matches this behavior with the linear one at early times $G^{(2)}_{\delta_L}(\bx_1,\bx_2,\t)\sim \tau^4/x^{3+n}_{12}$, one finds $z=4/(n+3)$. This means that the power spectrum in momentum space $\Delta_\delta(k,\t)=(k^3/2\pi^2)P_\delta(k,\t)$  at any order in perturbation theory must be a function of 
 $\Delta_\delta(k,\t)=\Delta_\delta(k/k_*(\t))$, where $k_*(\t)\sim \tau^{-4/(n+3)}$ and it can be interpreted as the momentum scale at which perturbations become non-linear at any given time.

Another, and maybe more interesting,  consequence of the symmetries we have discussed in the $\Lambda$CDM model is to produce consistency relations 
which correlators should satisfy in the squeezed limit, that is in the case in which one of the modes is  a long wavelength mode and therefore it may be assumed to evolve in the linear regime. This allows to relate $(n+1)$-correlation functions containing a soft mode to the 
to $n$-point correlation functions of the short modes. 

Consider, for instance, the $n$-point correlation function
of short modes density contrasts $\Big< \delta_{\vk_1}\delta_{\vk_2}\cdots\delta_{\vk_n}\Big>$. 
The points are supposed to be contained in a sphere of radius  $R$ much smaller than the long 
wavelength mode of size $\sim 1/q$ and centered at the origin of the coordinates.  According to what we have 
discussed in section 3, the non-relativistic equation of motions are invariant under the generic transformation $\t'\rightarrow t$ and  $\bx\rightarrow \bx+{\vec{n}}(T(\tau))$. This means that we can generate a 
long wavelength mode for the velocity perturbation $\bv_L(\t,\vec{0})$ just by choosing properly
the vector ${\vec{n}}(\tau)$

\be
{\vec{n}}(\tau)=-\int^\tau\d\eta\, \bv_L(\eta, \vec{0}) +{\cal O}(qR v_L^2).
\label{long}
\ee
In other words, the correlator of the short wavelength modes in the background of the long wavelength mode perturbation should satisfy the relation 
\be
\Big<\delta(\t_1,\bx_1)\delta(\t_2,\bx_2)\cdots\delta(\t_n,\bx_n)\Big>_{v_L}=
\Big<\delta(\t_1',\bx_1')\delta(\t_2',\bx_2')\cdots\delta(\t_n',\bx_n')\Big>.
\ee
This is nothing else that the statement that the effect of a physical long wavelength velocity perturbation  
onto the short modes
should be  indistinguishable from the long wavelength mode velocity generated by the transformation 
$\delta x^i={n}^i(\tau)$.
In momentum space one therefore obtains

\be
\Big< \delta_{\vq}(\t)\delta_{\vk_1}(\t_1)\cdots\delta_{\vk_n}(\t_n)\Big>_{q\to 0}= 
\Big< \delta_{\vq}(\t)\Big<\delta_{\vk_1}(\t_1)\cdots\delta_{\vk_n}(\t_n)\Big>_{v_L}\Big>. 
\ee

The variation of the  $n$-point correlator under such a transformation is given by

\begin{align}
\delta_n \Big<\delta(\t_1,\bx_1)\cdots\delta(\t_n,\bx_n)\Big>&=
\int \frac{\d^3\vk_1}{(2\pi)^3}\cdots \frac{\d^3\vk_n}{(2\pi)^3} \Big<\delta_{\vk_1}(\t_1)
\cdots\delta_{\vk_n}(\t_n)\Big>\nonumber \\
&\times \sum_{a=1}^n\delta x^i_a (i k_a^i) e^{i(\vk_1\cdot \vx_1+\cdots\vk_n\cdot \vx_n)}\nonumber \\
&=\int \frac{\d^3\vk_1}{(2\pi)^3}\cdots \frac{\d^3\vk_n}{(2\pi)^3} \Big<\delta_{\vk_1}(\t_1)
\cdots\delta_{\vk_n}(\t_n)\Big>\nonumber \\
&\times \sum_{a=1}^n n^i(\tau_a) (i k_a^i) e^{i(\vk_1\cdot \vx_1+\cdots\vk_n\cdot \vx_n)}.
\end{align}
Then we find that 
\begin{eqnarray}
\Big< \delta_{\vq}(\t)\delta_{\vk_1}(\t_1)\cdots\delta_{\vk_n}(\t_n)\Big>_{q\to 0}&=&
\Big< \delta_{\vq}(\t)\Big<\delta_{\vk_1}(\t_1)\cdots\delta_{\vk_n}(\t_n)\Big>_{v_L}\Big> \nonumber \\
&=&i\sum_{a=1}^n\Big< \delta_{\vq}(\t) n^i(\tau_a)\Big>  k_a^i \Big<\delta_{\vk_1}(\t_1)
\cdots\delta_{\vk_n}(\t_n)\Big>.
\end{eqnarray}
As for the   $\Lambda$CDM model we have

\begin{eqnarray}
\int^\tau\d\eta\, \bv_{\vq}(\eta)&=&i\frac{q^i}{q^2}\int^\tau\d\eta\,{\cal H}\,f(\eta)\,\delta_{\vq}(\eta)=i\frac{q^i}{q^2}\int^\tau\d\eta\,{\cal H}\,
\frac{1}{{\cal H}}\frac{\d\ln D_+(\eta)}{\d\eta}\,\frac{D_+(\eta)}{D_+(\eta_{\rm in})}\delta_{\vq}(\eta_{\rm in})
=i\frac{q^i}{q^2}\delta_{\vq}(\tau),\nonumber\\
&&
\end{eqnarray}
where $D_+$ is the linear growth factor, we finally get

\be
\fbox{$\displaystyle
\Big< \delta_{\vq}(\t)\delta_{\vk_1}(\t_1)\cdots\delta_{\vk_n}(\t_n)\Big>'_{q\to 0}
=- P_{\delta_L}(q,\tau)\sum_{a=1}^n \frac{D_+(\t_a)}{D_+(\tau)} \frac{{\vec q} \cdot \bk_a}{q^2}\Big<\delta_{\vk_1}(\t_1)
\cdots\delta_{\vk_n}(\t_n)\Big>'$},
\label{deltadelta}
\ee
where the primes indicate that one should remove the Dirac delta's coming from the momentum conservation and $P_{\delta_L}(q,\tau)=(D_+(\tau)/D_+(\tau_{\rm min}))^2P_\delta(q,\tau_{\rm in})$ is the linear matter power spectrum. Of course, similar consistency relations may be found involving the other quantities, the velocity perturbation and the gravitational potentials, in various combinations.
Notice that, if  the correlators are computed all at equal times, the right-hand side of eq. (\ref{deltadelta}) vanishes by momentum conservation and the $1/q^2$ infrared divergence will not appear when calculating invariant quantities. 
For the three-point correlator, we obtain

\be
\Big< \delta_{\vq}(\t)\delta_{\vk_1}(\t_1)\delta_{\vk_2}(\t_2)\Big>'_{q\to 0}
= -P_\delta(q,\tau) \left(\frac{D_+(\t_1)}{D_+(\tau)}-\frac{D_+(\t_2)}{D_+(\tau)}\right) \frac{{\vec q} \cdot \bk_1}{q^2}\Big<\delta_{\vk_1}(\t_1)\delta_{\vk_2}(\t_2)\Big>'.
\label{deltadeltadelta}
\ee 
One can easily check this result holds at second-order in perturbation theory in the matter-dominated era when \cite{sb}

\be
\delta^{(2)}_{\vk}(\tau)=\int\frac{\d^3 k_1}{(2\pi)^3}\frac{\d^3 k_2}{(2\pi)^3}\left[\frac{5}{7} +\frac{1}{2}(\vk_1\cdot\vk_2)\frac{k_1^2+k_2^2}{k_1^2 k_2^2}+\frac{2}{7}\frac{(\vk_1\cdot\vk_2)^2}{k_1^2 k_2^2}
\right]\delta^{(3)}(\vk-\vk_1-\vk_2)\,\delta^{(1)}_{\vk_1}(\tau)\delta^{(1)}_{\vk_2}(\tau).
\ee
Indeed, in the squeezed limit 

\be
\Big< \delta_{\vq}(\t)\delta_{\vk_1}(\t_1)\delta_{\vk_2}(\t_2)\Big>_{q\to 0}\simeq \Big< \delta^{(1)}_{\vq}(\t)\delta^{(2)}_{\vk_1}(\t_1)\delta^{(1)}_{\vk_2}(\t_2)\Big>_{q\to 0}+\Big< \delta^{(1)}_{\vq}(\t)\delta^{(1)}_{\vk_1}(\t_1)\delta^{(2)}_{\vk_2}(\t_2)\Big>_{q\to 0},
\label{aaa}
\ee
and recalling that $\delta_{\vk}(\t)=(D_+(\tau)/D_+(\tau_{\rm min}))\delta_{\vq}(\tau_{\rm in})$ with $D_+(\tau)=a(\tau)$, one recovers (\ref{deltadeltadelta}) by accounting
for the appropriate permutations and taking the leading middle term in the squared parenthesis of  Eq. (\ref{aaa}). 

Let us close with some comments. 
The consistency relation (\ref{deltadelta}) is  true at any order in perturbation theory.  As such, it  might represent a useful too to check  the findings of the 
various schemes dealing analytically with the problem of structure formation beyond the standard perturbation theory. Nevertheless, 
the consistency relation might be of more practical use and tested in  future galaxy surveys which are divided  into multiple redshift bins. Indeed, cosmic tomography makes it possible to map out the three-dimensional  distribution of mass and thus to observe correlators at different epochs. 
Of course, one needs the galaxy correlators and not the underlying dark matter ones. However, galaxies, once formed, obey the following equations on sub-Hubble scales

 \be
 &&\frac{\partial \delta_{\rm g}(\bx,\t)}{\partial \t}+\bnabla(1+\delta_{\rm g}(\bx,\t))\bv_{\rm g}(\bx,\t)=0\label{fg1},\\
 && \frac{\partial \bv_{\rm g}(\bx,\t)}{\partial \t}+{\cal{H}}(\t)\bv_{\rm g}(\vx,\t)
 +[\bv_{\rm g}(\vx,\t)\cdot \bnabla]\bv_{\rm g}(\bx,\t)=\-\bnabla\Phi(\bx,\t),\label{fg2}\\
 &&
 \nabla^2\Phi(\bx,\t)=\frac{3}{2} \Omega{\cal{H}}^2(\tau)\delta(\bx,\tau), \label{fg3}
 \ee
where $\delta_{\rm g}(\bx,\t)$ and $\bv_{\rm g}(\bx,\t)$ are the galaxy overdensity and peculiar velocity, respectively, while $\delta(\bx,\t)$
is the underlying dark matter overdensity. Following the same steps in sections 2 and 3, one can show that the set of equations (\ref{fg1}-\ref{fg3}) are invariant under the transformations 
\begin{align}
\delta_{\rm g}'(\bx,\t)&=\delta_{\rm g}(\bx',\t'),\label{gg1}\\
\bv_{\rm g}'(\bx,\t)&=\bv_{\rm g}(\bx',\t')-\dot{\vec{n}}(T),\label{gg2}\\
\Phi'(\bx,\t)&=\Phi(\bx',\t')+\Big{(}{\cal{H}}\dot{\vec{n}}(T)+\ddot{\vec{n}}(T)\Big{)}\cdot \bx.\label{gg3}
\end{align}
This is true even if the we do not assume $\bv_{\rm g}(\bx,\t)=\bv(\bx,\t)$,  that is  the galaxy peculiar velocity is unbiased,  as is often done.
Therefore the consistency relation (\ref{deltadelta}) should be true also for the galaxy overdensities, independently of the bias between 
$\delta_{\rm g}(\bx,\t)$ and $\delta(\bx,\t)$. 

Finally, it is  possible that gravity at large distances is modified by, for instance,
 a Yukawa-like modification of the Poisson equation

\be
(\nabla^2-a^2m^2)\Phi(\bx,\t)=\frac{3}{2} \Omega{\cal{H}}^2(\tau)\delta(\bx,\tau), \label{mass}
\ee
where $1/m$ defines some new infrared scale where gravity gets modified (the factor  $a^2$ in front of $m^2$ is 
such that the Yukawa correction becomes
small in the early universe, but  some other time
dependence is possible). In such a case, the symmetries 
(\ref{g1}-\ref{g3}) fail in general due to the mass term. The only case   (\ref{mass}) could  be
invariant is to choose a specific $\vec{n}(T)$ which will leave $\Phi$ itself invariant. 
This possibility is provided by 
\be
\vec{n}(\t)=\vec{n}_0+\vec{n}_1\int^\t\frac{\d\eta}{a(\eta)},
\ee
where $\vec{n}_0,\vec{n}_1$ are constant vectors. 
In this case, the induced transformation for the velocity $\vec{v}(\t,\bx)$ field
will be
\be
\vec{v}'(\vx,\t)=\vec{v}(\vx',\t')+\frac{1}{a(\tau)}\vec{n}_1,  
\ee
corresponding to a decaying mode. As the latter is of limited importance, we can safely  say that the symmetries 
we have discussed in this paper do not hold any longer and a violation
of the consistency relations might be a signal of modification of gravity.

\noindent 
 \section*{Acknowledgments}
A.R. is supported by the Swiss National
Science Foundation (SNSF), project `The non-Gaussian Universe" (project number: 200021140236).

 \section*{Note added}
When completing this work, we became aware of a similar work by M. Peloso and M. Pietroni. Our results, when overlap is possible, agree with theirs. We thank them for useful discussions. 


\begin{thebibliography}{99}

\bibitem{coleman} See, for instance, `Ëspects of Symmetry: Selected Erice Lectures" by S. Coleman, Cambridge University press (1988). 

\bibitem{WT} Y. Takahashi, Nuovo Cimento, Ser. 10, {\bf 6} (1957) 370; J.C. Ward, Phys. Rev. {\bf 78}, (1950) 182.

\bibitem{lrreview} D.~H.~Lyth and A.~Riotto,
  Phys.\ Rept.\  {\bf 314}, 1 (1999)
  [hep-ph/9807278].
  
  
  \bibitem{antoniadis} 
  I.~Antoniadis, P.~O.~Mazur and E.~Mottola,
  JCAP {\bf 1209}, 024 (2012)
  [arXiv:1103.4164 [gr-qc]].
  
  
  \bibitem{creminelli1} 
  P.~Creminelli,
  Phys.\ Rev.\ D {\bf 85}, 041302 (2012)
  [arXiv:1108.0874 [hep-th]].
  
  
  \bibitem{us1} 
  A.~Kehagias and A.~Riotto,
  Nucl.\ Phys.\ B {\bf 864}, 492 (2012)
  [arXiv:1205.1523 [hep-th]].
  
  
  \bibitem{us2} 
  A.~Kehagias and A.~Riotto,
  Nucl.\ Phys.\ B {\bf 868}, 577 (2013)
  [arXiv:1210.1918 [hep-th]].
  
  
  
  \bibitem{creminelli2} 
  P.~Creminelli, J.~Norena and M.~Simonovic,
  JCAP {\bf 1207}, 052 (2012)
  [arXiv:1203.4595 [hep-th]].
  

\bibitem{hui} 
  K.~Hinterbichler, L.~Hui and J.~Khoury,
  JCAP {\bf 1208}, 017 (2012)
  [arXiv:1203.6351 [hep-th]].

  

  
  \bibitem{baumann1} 
  V.~Assassi, D.~Baumann and D.~Green,
  JCAP {\bf 1211}, 047 (2012)
  [arXiv:1204.4207 [hep-th]].
  
 

\bibitem{baumann2} 
  V.~Assassi, D.~Baumann and D.~Green,
  arXiv:1210.7792 [hep-th].


\bibitem{mal} 
  J.~M.~Maldacena,
  JHEP {\bf 0305}, 013 (2003)
  [astro-ph/0210603].

\bibitem{cz} 
  P.~Creminelli and M.~Zaldarriaga,
  JCAP {\bf 0410}, 006 (2004)
  [astro-ph/0407059].


\bibitem{sb} For a review, see 
  F.~Bernardeau, S.~Colombi, E.~Gaztanaga and R.~Scoccimarro,
  Phys.\ Rept.\  {\bf 367}, 1 (2002)
  [astro-ph/0112551].

\bibitem{rpt} 
  M.~Crocce and R.~Scoccimarro,
  Phys.\ Rev.\ D {\bf 73}, 063519 (2006)
  [astro-ph/0509418].

\bibitem{rg} 
  S.~Matarrese and M.~Pietroni,
  JCAP {\bf 0706}, 026 (2007)
  [astro-ph/0703563].


\bibitem{val} 
  P.~Valageas,
  Astron.\ Astrophys.\  {\bf 421}, 23 (2004)
  [astro-ph/0307008].

\bibitem{mat} 
  T.~Matsubara,
  Phys.\ Rev.\ D {\bf 77}, 063530 (2008)
  [arXiv:0711.2521 [astro-ph]].

\bibitem{rb}  R.~Brustein and A.~Riotto,
  JCAP {\bf 1111}, 006 (2011)
  [arXiv:1105.4411 [astro-ph.CO]].

\bibitem{effective} 
  J.~J.~M.~Carrasco, M.~P.~Hertzberg and L.~Senatore,
  JHEP {\bf 1209}, 082 (2012)
  [arXiv:1206.2926 [astro-ph.CO]].






\bibitem{jb}  B.~Jain and E.~Bertschinger,
  Astrophys.\ J.\  {\bf 456}, 43 (1996)
  [astro-ph/9503025].
  
\bibitem{scocgal} R.~Scoccimarro and J.~Frieman,
  Astrophys.\ J.\ Suppl.\  {\bf 105}, 37 (1996)
  [astro-ph/9509047].

\bibitem{scocself} R.~Scoccimarro and J.~Frieman,
  Astrophys.\ J.\  {\bf 473}, 620 (1996)
  [astro-ph/9602070].

\bibitem{pd} M. Davis and  P.J.E. Peebles, Astrpphys. Journal  {\bf 34} (1977), 425.

\bibitem{Duval} 
  C.~Duval, G.~Burdet, H.~P.~Kunzle and M.~Perrin,
  Phys.\ Rev.\ D {\bf 31}, 1841 (1985).
  \bibitem{pig} 
  E.~Prugovecki,
  Class.\ Quant.\ Grav.\  {\bf 4}, 1659 (1987).

\bibitem{gibbons}
C.~Duval, G.~W.~Gibbons and P.~Horvathy,
  Phys.\ Rev.\ D {\bf 43}, 3907 (1991)
  [hep-th/0512188].
  \bibitem{horvathy}
   P.~A.~Horvathy and P.~-M.~Zhang,
  Eur.\ Phys.\ J.\ C {\bf 65}, 607 (2010)
  [arXiv:0906.3594 [physics.flu-dyn]].



 \end{thebibliography}
\end{document}